# Dispersing bimagnons and doping induced bimagnon-charge modes in superconducting cuprates

L. Braicovich[1], L.J.P. Ament[2], V. Bisogni[3], F. Forte[4], C. Aruta[5], G. Balestrino[5], N.B. Brookes[3], G. M. De Luca[6], P.G. Medaglia[5], F. Miletto Granozio[6], M. Radovic[6], M. Salluzzo[6], J. van den Brink[7], and G.Ghiringhelli[8*]

[1] INFM-CNR-SOFT, and Dipartimento di Fisica, Politecnico di Milano, Italy

[2] Institute-Lorentz for Theoretical Physics, Universiteit Leiden, The Netherlands

[3] European Synchrotron Radiation Facility, Grenoble, France

[4] Dipartimento di Fisica "E. R. Caianiello", Università di Salerno, and INFM-CNR Laboratorio Regionale SuperMat, Italy

[5] CNR-INFM-COHERENTIA, and Dipartimento di Ingegneria Meccanica, Università di Roma Tor Vergata, Italy

[6] CNR-INFM COHERENTIA, and Dipartimento di Scienze Fisiche, Università di Napoli "Federico II", Italy

[7] Institute for Molecules and Materials, Radboud Universiteit Nijmegen, The Netherlands

[8] INFM-CNR-COHERENTIA, and Dipartimento di Fisica, Politecnico di Milano, Italy

**In the early days of high temperature superconductivity it was already recognized that magnetic properties of these materials are intimately related to the superconducting ones[1]. When doped, the long-range ordered antiferromagnetic background of pristine copper-oxide insulators melts away and makes room for a spin liquid and superconductivity. By resonant inelastic x-ray scattering (RIXS)[2] in the soft regime we probe the hitherto inaccessible dynamical multiple-spin correlations of the magnetic background in a series of parent compounds and in high $T_c$ materials [NCCO ($Nd_{2-x}Ce_xCuO_4$) and LSCO ($La_{2-x}Sr_xCuO_4$)]. High resolution measurements allows the clear observation of dispersing bimagnon excitations. In the undoped compounds the theory[3,4] fits the data on these coherent spin excitations without free parameters. In nearly optimally doped LSCO we observe the appearance of a new collective excitation at an energy of 250 ± 60 meV having the signature of a coupled bimagnon-charge mode. It has a strongly reduced dispersion and lies in a so far unexplored region of momentum and energy space in the mid-infrared.**

Up to now the spin dynamics of the high $T_c$ cuprates have been studied with experimental methods (mostly neutron scattering[5,6,7]) that provide momentum resolved information on single spin-flip processes, often called magnons. Resonant inelastic x-ray scattering (RIXS) is qualitatively different because it probes dynamics of two spin-flips[3] and thus provides a means to observe bimagnon excitations, i.e., coherent states of two magnons[8]. The unique feature of RIXS is that it allows the measurement of the bimagnons *dispersion* by determining both momentum change and energy loss of the scattered x-ray photons. Such is far beyond the capabilities of traditional low energy optical techniques[9,10,11,12,13], which are constrained to zero momentum transfer because, as opposed to high energy x-rays, photons in the visible range carry negligible momentum.

Copper $L_3$ edge RIXS of cuprates begins with the absorption of an x-ray photon, resonantly promoting an electron from the inner Cu $2p$ state to its $3d$ valence shell. As in the pristine cuprates Cu is basically in a $3d^9$ configuration, photon absorption (around 931 eV) corresponds to producing a $2p^63d^9 \rightarrow 2p^53d^{10}$ exciton. A unique feature of RIXS is that we can select this particular excitation channel also for the doped cuprates, even if in this situation the additional holes in the system provide a number of alternative absorption channels. This is obtained by tuning to the $2p^63d^9 \rightarrow 2p^53d^{10}$ main peak[14]. The second step is the decay of the intermediate state, $2p^53d^{10} \rightarrow 2p^63d^{9*}$, which behind leaves the material in an excited state indicated by the *, and produces outgoing photons of which we measure energy and momentum. Whereas in the initial and final state of the system a spin 1/2 is present in the copper $3d$ shell, in the $3d^{10}$ intermediate state the $3d$ shell is completely filled, having no spin at all. So in Cu $L_3$ edge RIXS a non-magnetic impurity is introduced in the intermediate state. This impurity is dynamically screened by the rest of the system, inducing spin reorientations at sites around it, of which the lowest order allowed ones are bimagnon excitations (Fig. 1). Besides the spin excitations all RIXS spectra show intense *d-d* excitations characterized by the redistribution of the electrons within the $3d$ shell.

For $CaCuO_2$ (CCO), $La_2CuO_4$ (LCO), LSCO and NCCO, Figs. 2b, 2c, 2d show, with unprecedented energy resolution ($\Delta E$ = 400 meV), the scattered intensity versus the photon energy loss[15,16,17]. Figs. 2e, 2f show how the spectra change while samples are rotated, so to change $q$, i.e., the projection onto the *ab* crystallographic plane of the momentum transferred from the scattering photon to the sample. To the authors' knowledge these $q$-dependent data are the first of its kind. A low energy loss feature is clearly discerned, having for large $q$ an energy up to about 400 meV. The first key result is that this feature is ubiquitous, going from undoped cuprates (CCO, LCO and NCO ($Nd_2CuO_4$)) to hole (LSCO) or electrons (NCCO) doped materials, for any incident photon polarizations. The second key result is its strong dispersion with the sample orientation. For two undoped compounds with different crystal structures (CCO and LCO) and for one superconductor (LSCO) we have mapped this dispersion out in detail [see Fig. 3].

First we present the experimental evidence for the absence of *d-d* excitations in the low energy feature. It is known[16] that the strong peaks in RIXS around 2 eV energy loss are due to *d-d* excitations. The energy of the *d-d* excitations changes rather drastically with crystal structure, as one can expect from the large differences in crystal field splitting of materials with apical oxygen (LCO and LSCO) and without it (CCO, NCO and NCCO), as shown in Figs. 2b, 2c, 2d. In contrast, the low energy feature energy does not depend on the presence of apical oxygen atoms, thus excluding it being a *d-d* excitation. Further experimental evidence is provided by the *q* dependence shown in Figs. 2e, 2f, for CCO at various angles of incidence to the surface ($\theta = 55°$ is the specular geometry corresponding to *q* = 0, while equal and opposite values of *q* correspond to angles symmetric with respect to 55°). The low energy feature in the raw data has a dispersion that is basically symmetric around 55°, which is indicative of the wave-like nature of this excitation. On the other hand the *d-d* excitations around 2 eV evolve monotonically from 10° to 90°, without any symmetry around *q* = 0.

A magnetic origin of this feature, instead, is in accordance with our measurements, as the strength of the magnetic exchange interaction in these different materials is similar. Another strong indication for a magnetic origin is the dependence of the RIXS spectra on incoming photon polarization. The first step of the RIXS process, the x-ray absorption, depends on photon polarization[18]. And the *d-d* part of the spectrum is very sensitive to the polarization at all angles: both shape and intensity change, as shown in reference 16. On the contrary the low energy feature is polarization independent, within the experimental accuracy, when the spectra for the two polarizations are normalized to the total *d-d* peak intensity (see Figs. 2b, 2c). This is precisely what we expect for magnetic scattering caused by the non-magnetic $d^{10}$ intermediate state, because the photon polarization does not change the character of the closed shell $d^{10}$ configuration: it only differentiates between different possible orbital characters of the 2p core hole.

That the magnetic excitation rather has bimagnon than single magnon character becomes clear from its dispersion in LCO shown in Fig. 3b. Along the dispersion curve, our measured energies are systematically and significantly higher than those of single magnon measured with neutrons[Error! Bookmark not defined.]. They are therefore incompatible with a single magnon interpretation: indeed, from theory, we know that there is only a very small single-magnon contribution submerged in the bimagnon signal. The assignment to bimagnon is also supported by the interpretation of very recent Cu K edge RIXS data[19] on LCO. Also in the K edge experiment a loss feature around 500 meV was observed and attributed to bimagnons, even if its dispersion could not be determined and the authors do not exclude there a *d-d* contribution.

Having established empirically the bimagnon origin of the dispersing low energy feature, we bolster the case by comparing the data to theory. In the context of K edge RIXS on Mott insulators we have previously developed the microscopic theory [3,4] for multiple-magnon scattering that results from the sudden change of the magnetic moment in the intermediate state, relying on the so-called ultrashort core-hole lifetime expansion[20]. It follows from elementary considerations that the theoretical descriptions of resonant scattering at the L and K edge are equivalent if we stop the expansion at the first order. This allows a direct comparison of the experimental bimagnon dispersion to the theoretical one shown in Fig 3. The agreement is excellent both for CCO and LCO, in particular when considering that our theory contains only one parameter, the antiferromagnetic nearest neighbour exchange coupling *J*, which we take from neutron measurements (LCO[7,21]), so that the fit contains no free parameters. In CCO our fitting gives the same *J* within the error bars. From a theory viewpoint there are two major differences between magnetic RIXS at the L and K edge. At the L edge the core hole lifetime is longer and, at the same time, the $3d^{10}$ intermediate configuration produces a stronger magnetic perturbation because it completely blocks all magnetic exchange paths involving the site with the core hole. These two factors work together in enhancing scattering terms given at the second order expansion; this higher order contribution should become visible at $q = 0$ where the first order term produces no scattering. This $q = 0$ second order bimagnon contribution is computed to have a very broad maximum around $E = 4.2 J$, which is indeed what we observe in this energy region (black square in Fig. 3).

The spin liquid phase of the doped superconductor LSCO shows a dramatic reconstruction of the low energy dispersion, see Fig. 3c. Particularly at zero transferred momentum, $q = 0$, the spectral weight now piles up around $250 \pm 60$ meV. Due to the presence of low-energy charge degrees of freedom first order RIXS scattering is not anymore forbidden at $q = 0$. This feature is the signature of a mixed charge-bimagnon mode. At $q = 0$ the charge-bimagnon mode is of $A_{1g}$ symmetry, which indicates that it is intimately related to the superconducting state. It is in the superconducting phase only that optical Raman experiments have previously been able to uncover a mode of this symmetry, having in this case an energy of around 40 meV -- the scale of the superconducting gap[22]. This massive two-magnon excitation[23] has been interpreted as the superconducting Anderson-Bogoliubov amplitude mode[24,25]. This suggests that the higher energy $A_{1g}$ mode that we observe is the incoherent spectral weight of the quasiparticles that, upon gaining coherence at lower temperatures, condense into the superconducting state. Testing this conjecture and probing the dispersion of these quasiparticles at the superconducting energy scale asks for future RIXS work with a very high resolving power, $\Delta E < 30$ meV, i.e., $E/\Delta E > 30000$ at the Cu $L_3$ edge. Although this is well beyond recent achievements[26] the current progress in instrumentation is such that this tantalizing possibility can be achieved within a few years.

Finally we stress that soft x-ray RIXS offers new possibilities in the measurement of magnetic excitations, cross fertilizing with neutrons and expanding dramatically the experimental possibilities. In fact RIXS can be equally measured on single crystals and thin film, whereas neutron inelastic scattering require massive samples not always available. Moreover RIXS can almost uniquely excite the coupled spin-charge modes as we demonstrated here for doped cuprates.

**Acknowledgements**

The authors are deeply indebted to G.A. Sawatzky for illuminating discussions and encouragement and to M. van Veenendaal and M. Grioni for very stimulating exchange of ideas. The authors thank also L.H. Tjeng for making the NCO and NCCO samples available. The measurements have been taken under the AXES (Advanced X-ray Emission Spectroscopy) contract between the ESRF-Grenoble and the INFM/CNR of Italy.

**FIGURES 1-3**

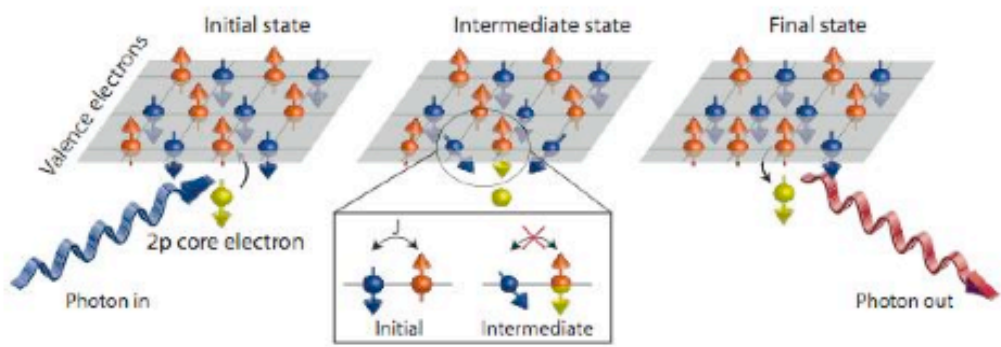

Figure 1

**Figure 1**. Schematic representation of the different steps in the resonant inelastic x-ray scattering (RIXS) process creating a bimagnon. By the absorption of the incident photon by one copper atom a 2$p$ core electron is promoted into the 3$d$ shell. Since this fills up the 3$d$ shell to 3$d^{10}$, the excited ion acts as a non-magnetic impurity, which is screened by the surrounding spins as long as the intermediate state lives. Upon de-excitation and emission of the outgoing photon the system can be left in an excited final state differing from the initial state by the presence of two magnon in the neighborhood of the scattering site. For readability the spins are sketched as perpendicular to the CuO$_2$ plane, but the argument applies to any spin orientations in the 2D antiferromagnetic squared spin lattice.

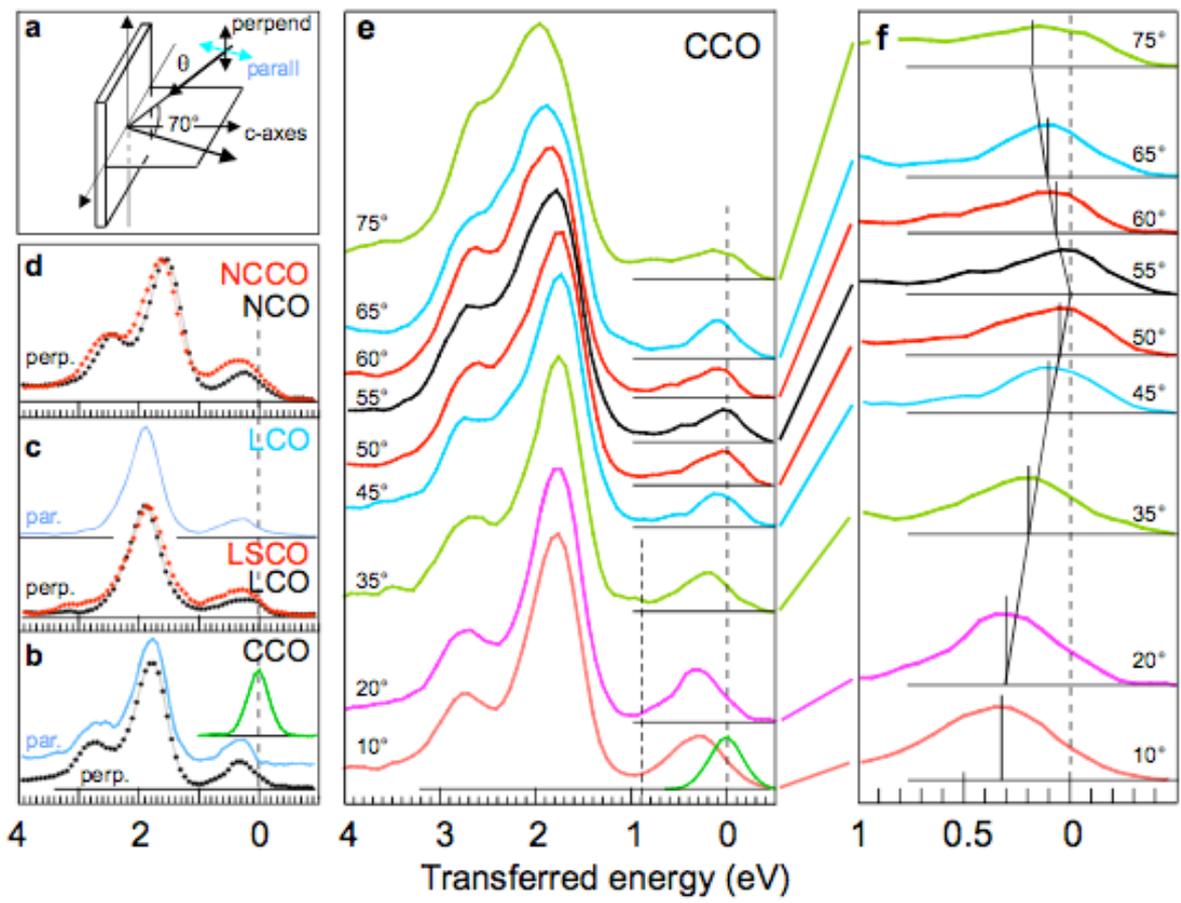

Figure 2

**Figure 2**. **Experimental setup and examples of raw RIXS spectra.** Panel a: schematics of the experimental geometry and definition, with respect to the scattering plane, of the incident photon polarization state. Panels b, c, d: Cu $L_3$ (main absorption peak) RIXS spectra, incidence at $\theta = 20°$ to the sample surface, which was in all cases perpendicular to the *c* axis. Panel b: CCO, polarization perpendicular (black) and parallel (blue) to the scattering plane (spectra normalized to the same main peak intensity). The overall instrumentation response is given by the non resonant elastic peak (green line), measured on Ag paint at the same incident photon energy used for the RIXS spectra: the full width at half maximum is 400 meV. Panel c: LCO (black) and LSCO (red) with perpendicular polarization, LCO (blue) with parallel polarization (spectra normalized to the same main peak intensity). Panel d: NCO and NCCO. Panels e, f: CCO, various incidence angles $\theta$ to the surface. To highlight the dispersion of the low energy feature, in panel f the spectra are vertically offset proportionally to *q*, the component parallel to the *ab* plane of the moment transferred from the scattering photon to the sample. The spectrum with *q* = 0 ($\theta = 55°$) is in black while the same colour is used for the spectra having same |*q*| but opposite directions.

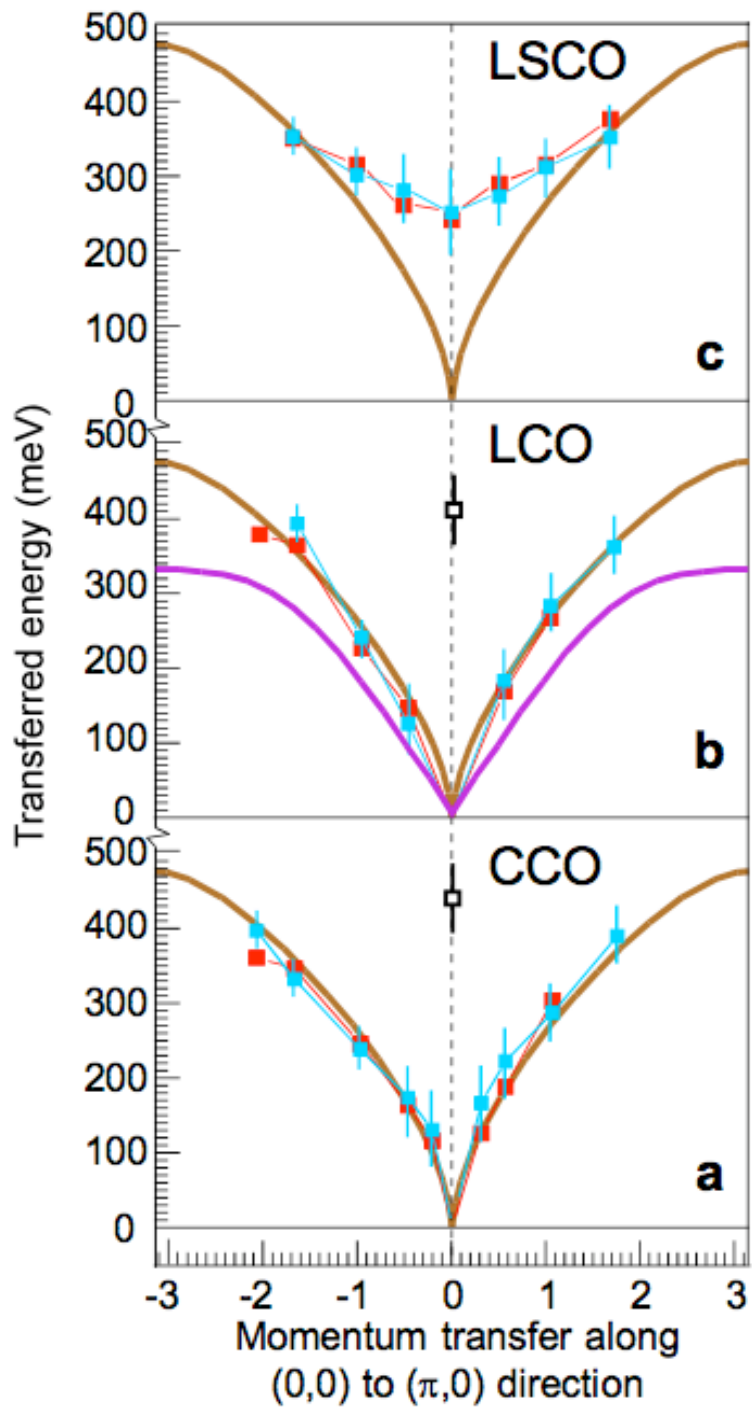

Figure 3

**Figure 3. Experiment and theory: dispersion of bimagnons in undoped and doped cuprates.** The peak energy position of the low energy feature is shown as function of $q$. The experimental results are given by the data points (perpendicular polarization in red and parallel polarization in blue). For readability we give only the error bars with parallel polarization. The error bars come from the uncertainty of elastic contribution subtraction and of selfabsorption/saturation corrections. As the angle θ increases towards normal incidence (conventionally positive $q$) the selfabsorption correction becomes more important and the error bars increase. Around $q = 0$ the error bars are greater due to the difficulty in evaluating small energy losses. In panel b we give also the one magnon dispersion (in purple) taken from Ref. 7. The theoretical dispersion in parent compounds including second and third neighbours and accounting for experimental resolution is given by the heavy brown lines (panels a and b). In this theory the exchange coupling $J = 120$ meV in LCO is taken from room temperature neutron measurements[7] with renormalization as in Ref. 21 In panel c (LSCO) the thin brown line is the theoretical dispersion in LCO given for comparison.

---